# Observations of Magnetospheric Solar Wind Charge Exchange

R. Ringuette[1,2], K. D. Kuntz[2,3], D. Koutroumpa[4], P. Kaaret[5,6], D. LaRocca[5], J. Richardson[5,7]

[1]ADNET Systems, Inc., 6720 B Rockledge Dr., Suite 504, Bethesda, MD 20817, USA; rebecca.ringuette@gmail.com

[2]NASA Goddard Space Flight Center, Greenbelt, MD 20771, USA

[3]The Henry A. Rowland Department of Physics and Astronomy, Johns Hopkins University, Baltimore, MD 21218, USA

[4]LATMOS/IPSL, CNRS, UVSQ Paris-Saclay, Sorbonne Université, Guyancourt, France

[5]Department of Physics and Astronomy, University of Iowa, Van Allen Hall, Iowa City, IA 52242, USA

[6]NASA Marshall Space Flight Center, Huntsville, AL 35812, USA

[7]Earth Resources and Observation Science Center, United States Geological Survey, 47914 252nd St., Sioux Falls, SD, 57198, USA

**Abstract**

The study of solar wind charge exchange (SWCX) emission is vital to both the X-ray astrophysics and heliophysics communities. SWCX emission contaminates all astrophysical observations in X-rays regardless of the direction. Ignoring this contribution to X-ray spectra can lead to erroneous conclusions regarding the astrophysical plasmas along the line of sight due to the similar spectral distributions of SWCX and several common types of more distant astrophysical plasmas. Since its discovery, literature has distinguished between diffuse SWCX emission resulting from solar wind-neutral interactions within the Earth's magnetosphere, called magnetospheric SWCX, and similar interactions occurring more generally throughout the heliosphere, called heliospheric SWCX. Here, we build upon previous work validating a modeling method for the heliospheric SWCX contribution in X-ray spectra obtained with a medium resolution CubeSat instrument named *HaloSat* at low ecliptic latitudes. We now apply this model to a specially designed set of extended observations with the same instrument and successfully separate the spectral contributions of the astrophysical background and the heliospheric SWCX from the remaining contributions. Specifically, we find significant excess emission for four observations in the O VII emission line not explained by other sources, possibly indicative of magnetospheric SWCX. We discuss these results in comparison with simulation results publicly available through the Community Coordinated Modeling Center. We also report an absorbed high-temperature component in two of the twelve fields of view analyzed.

## 1. Introduction

Solar wind charge exchange (SWCX) emission was discovered when comets proved to be strong emitters in the ROSAT soft X-ray energy bands (Lisse, 1996). The emission is produced when highly charged ions in the solar wind (e.g. O7+) capture an electron from nearby neutrals (Cravens, 1997). The newly created ions are in an excited state and emit soft X-ray photons (< 2 keV) as they de-excite into the ground state. Besides comets, SWCX emission is produced in planetary exospheres, the Earth's magnetosheath and the heliosphere. The latter two components (named m-SWCX and h-SWCX from now on) are a challenge for X-ray astrophysicists, since they contribute a variable foreground with similar spectral characteristics as some astrophysical thermal plasmas (e.g. Local Hot Bubble, galactic halo) for CCD-like spectral resolution (Cox, 1998; Kuntz, 2018).

The most conspicuous signature of SWCX ever observed was the excess emission, named "Long Term Enhancements - LTEs", in the ROSAT All-Sky Survey (RASS) light curves and uncleaned maps of the ¼ keV band (e.g., Snowden et al. 1994, 2009). Cravens et al. (2001) first modeled the LTEs and demonstrated the correlation of their signal variation with solar wind flux variations using both m-SWCX and h-SWCX contributions.

H-SWCX is produced by charge exchange with interstellar neutrals, hydrogen and helium, flowing through the heliosphere into interplanetary space. The two species have different density distributions due to the Sun's effects (Figure 1). Hydrogen is strongly ionized through proton charge exchange and subject to radiation pressure, which excludes the atoms from an area of 1-2 AU around the Sun (e.g. Quémerais et al. 1999 and the left panel of Figure 1). For helium, the ionization is much lower (mostly produced by EUV photons) and the radiation pressure is too weak to counterbalance solar gravitation, which forms an enhanced neutral density region (the He focusing cone) in the downstream direction of the Sun (e.g. Dalaudier et al. 1984, and the right panel of Figure 1). Both density distributions are roughly axisymmetric around the interstellar flow velocity vector, and closely reflect the changes of the 11-year solar activity cycle. Thus, the main h-SWCX signatures are solar cycle scale variations (Qu et al. 2022) and spatial effects due to variations in neutral column densities (Koutroumpa et al. 2009; Galeazzi et al. 2014). Short scale temporal variations due to the solar wind intrinsic variability tend to be smoothed out due to the large emissive volume of the heliosphere.

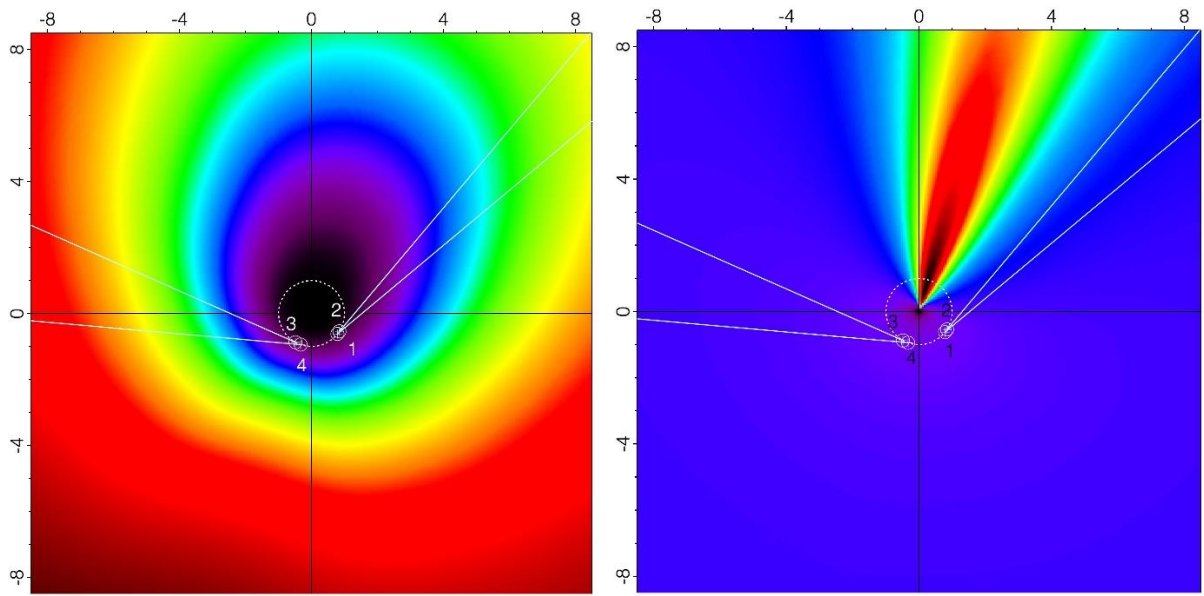

**Figure 1:** The densities of Hydrogen (left) and Helium (right) in the heliosphere centered on the Sun in arbitrary units. Reds indicate higher densities while blues and purples indicate lower densities with the axes values given in AU. The dashed circle in the middle of each panel indicates the Earth's orbit around the Sun. The numbers along the circles correspond to the observations marked with an asterisk in Table 4, and the thin white lines indicate the look directions of those observations. The cone-shaped feature in the right panel is called the He-focusing cone.

M-SWCX is produced from collisions with neutrals in the Earth's outer hydrogen layer of the atmosphere, spreading from ~10 R_E in the subsolar region and up to the Moon's orbit in the night-side (Baliukin et al. 2019). The Earth's magnetopause generally prevents the solar wind from interacting with the denser neutral layers, so the emission is mostly limited to the subsolar and flank portions of the magnetosheath and the polar cusps where ions can penetrate more deeply through open magnetic field lines (e.g. Robertson et al. 2006). However, during explosive events, such as Coronal Mass Ejections, that induce geomagnetic storms, the magnetopause boundaries are pushed back exposing the denser layers of the geocorona to enhanced solar wind fluxes (e.g. Sun et al. 2019). Therefore, the m-SWCX signal is mostly recognizable through its short-time-scale variations due to extreme solar wind conditions, and is the predominant source of LTE variability.

The two components are generally indistinguishable from each other (and from diffuse galactic sources) without proper modeling, since they are observed simultaneously in any direction, except in the strictly anti-solar direction in the night side of the Earth where m-SWCX should be negligible. The only studies that managed to isolate the m-SWCX signal are Dark Moon observations with ROSAT (Scmitt et al. 1991) and Chandra (Wargelin et al., 2004), since more distant signals (heliospheric and galactic) are blocked by the Moon's body.

Many studies attempted to characterize the SWCX contamination in X-ray observations by targeting the highly variable m-SWCX. Several cases of enhanced m-SWCX (and h-SWCX) emission were observed with Chandra, XMM-Newton and Suzaku. The general approach consists in investigating the correlation between temporal or spectral changes in the measured signal and solar wind variability (e.g. Snowden et al. 2004; Kuntz and Snowden 2008; Carter et al. 2009, 2011; Ezoe et al. 2011; Henley and Shelton 2012; Ishikawa et al. 2013; Ishi et al. 2019). Some papers also include modeling of the m-SWCX variable component (Wargelin et al., 2014; Whittaker et al. 2016) and occasionally the h-SWCX more stable component as well (Slavin et al. 2013). Most of these cases were fortuitous observations of m-SWCX variations, and the multitude of different cases and ample discussions highlight the complexity of the problem and its parameters (observing geometry, solar wind conditions, etc.). One study was specifically designed to investigate the m-SWCX emission of the dayside magnetosheath, while viewing a dark area of galactic emission, and compare it to an m-SWCX emission model (Snowden et al. 2009). The long (~100ks) observation was performed during the 2008 solar minimum, and although the signal was faint, the correlation with the model was significant to better than 99.9%.

In this work, we detail our efforts to separate spectral contributions due to m-SWCX from those of heliospheric SWCX and the astrophysical background in the X-ray spectral regime. Our approach begins with a novel observation strategy designed to compare observations of the same field of view with looks directions both away from and through the emissive portions of the Earth's magnetosheath, thus minimizing the magnetospheric SWCX contributions in one set of spectra and maximizing the same in another. A similar observation approach is not as effective for heliospheric SWCX emission since the emission is present in non-negligible amounts in every direction. Instead, we rely on a previously validated method to calculate the spectral contribution due to heliospheric SWCX in each observation based on the observed time history of the solar wind, the line of sight for each observation, and a physical emission model (Ringuette et al. 2021). Consequently, we independently determine the amplitudes and shapes of the spectral components of the astrophysical background, heliospheric SWCX, and magnetospheric SWCX.

Section 2 describes our observation strategy, data preparation methods, and the resulting list of observations. Section 3 presents our spectral modeling procedure, both for the set of spectra taken away from the most emissive portions of the Earth's magnetosheath, called tail spectra, and spectra taken of the same celestial positions but through the flanks of the Earth's magnetosheath, called flank spectra. The results of our spectral analyses are detailed in Section 4, including detection of higher temperature components in the Galactic halo in two of our chosen fields of view. We report significant excess in O VII emission for four flank spectra, likely due to magnetospheric SWCX contributions, and we extend our analysis in Section 5 to compare the observed excesses with those predicted by a physical model and discuss the results there. Section 6 compares the spectral analyses results obtained here with those obtained in recent works. We present our conclusions in Section 7.

## 2. Observations
*2.1 Observing Strategy*

Our goal in this work is to separate the contributions of magnetospheric SWCX emission from those of heliospheric SWCX and the astrophysical background. This is particularly challenging due to the similar spectral shapes of the three types of contributions, specifically the indistinguishable spectral shapes of the two types of SWCX emission, and the warm-hot component of the Galactic halo in the astrophysical background of the same observations. We approach this challenge by first designing a novel two-pronged observation strategy. One set of observations are designed with lines of sight through the tail of the Earth's magnetosheath to minimize magnetospheric SWCX spectral contributions, with a second set of the same fields of view with lines of sight through the flanks of the Earth's magnetosheath.

Such an observation strategy demands a large amount of observation time which is generally difficult to obtain. Additionally, the SWCX emission we aim to detect is diffuse and faint, requiring a large grasp. The spectral range of the emission also requires the observation instrument to have at least moderate energy resolution in the lower energy X-rays. These requirements were well met by the *HaloSat* CubeSat mission (LaRocca et al. 2020, Zajczyk et al. 2020, Kaaret et al. 2019).

The *HaloSat* CubeSat mission was the first astrophysics CubeSat mission funded by NASA. The instrument collected moderate resolution spectra in the soft X-rays from low Earth orbit during the recent solar minimum for approximately two years (late 2018 to early 2021). The non-imaging instrument mapped diffuse emission from the O VII line across the entire celestial sphere (Kaaret et al. 2020). The instrument design requirements for *HaloSat* were specifically chosen for the study of the Galactic Halo, which also fit well with SWCX as its secondary science goal, specifically the instrument's 14° diameter non-imaging field of view for each of three nearly identical silicon drift detectors, the 0.4 – 7 keV energy range, and the 85 keV energy resolution at 676.8 eV. As part of the mission planning, twelve fields of view chosen to minimize bright sources in the 0.4 to 7 keV energy range and extensive observations of each were planned. Observations of each field of view were planned with lines of sight down the tail of the Earth's magnetosheath, where the expected magnetospheric SWCX contribution is negligible, and of the same fields of view but with lines of sight through the flanks of the Earth's magnetosheath (see Figure 2). The observation limitations of the instrument prevented lines of sight through the brighter portions of the Earth's magnetosheath on the day-side of the orbit, and the modeling limitations of our heliospheric SWCX model required the chosen fields of view to be near the ecliptic plane.

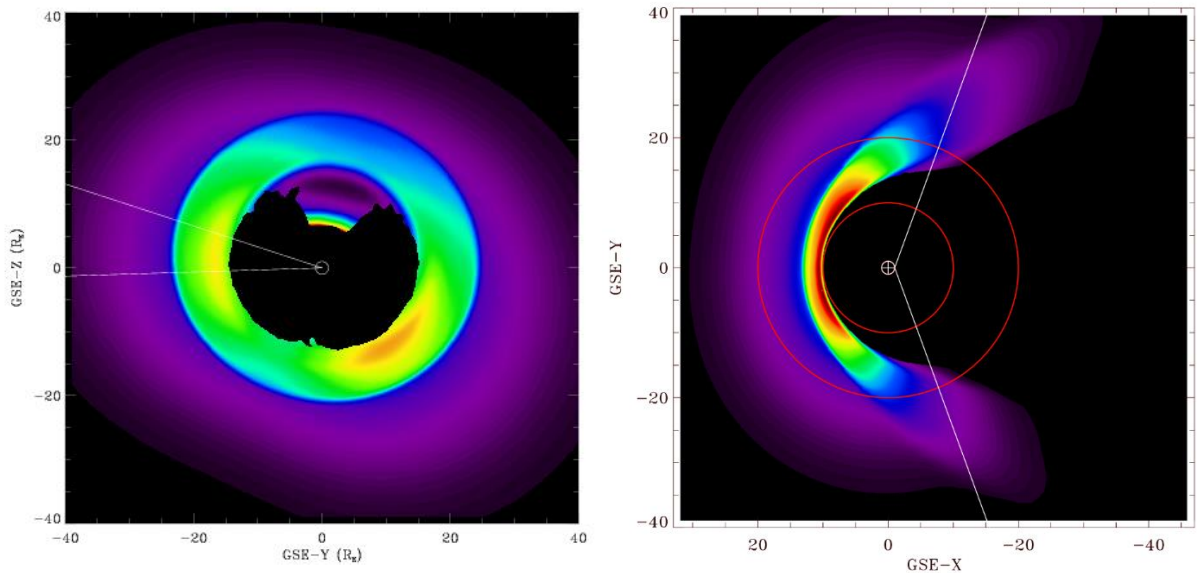

**Figure 2:** The X-ray emissivity in arbitrary units, calculated by the BATS-R-US magnetohydrodynamic model. *Left:* A slice in the GSE-X = 0 plane. The Earth is at the center of the plot and the Sun is behind you. A typical viewing geometry is shown for the HaloSat spacecraft at local midnight, with the viewing region bounded by the white lines. The dark region in the center of the plot is the area inside the magnetopause where the solar ions are, to first order, excluded. (The irregularity of this boundary is due to the field tracing algorithms and the step-size of the BATS-R-US model.) *Right:* A cut through the emissivity in the GSE-Z = 0 plane. The Earth is indicated by the crosshairs near the center of the plot and the Sun is to the left. The red rings indicate radial distances of 10 and 20 R_E. Sun angles of 110° are shown in white.

*2.2 Data Preparation*

As in Ringuette et al. (2021), the spectra analyzed in this work were also of fields of view chosen to minimize contribution from bright sources and use the same instrument. So, we take an almost identical approach to prepare the data for spectral analysis. The observed spectral components associated with the astrophysical background are constant on our observational timescales (compared to cosmic timescales), and the temporally varying SWCX contributions occur below about 2 keV. Therefore, any observed variation in the photon count rate at 3 keV and above is due to the 'particle' or 'instrumental' background associated with the instrument's changing location and other local factors.

To reduce this component in our spectra, we binned the count rates in the 3-7 keV and >7 keV energy ranges into 64-second chunks and remove sections with count rates above 0.12 counts $s^{-1}$ in the 3-7 keV energy range or above 0.75 counts $s^{-1}$ in the >7 keV energy range. The remaining sections of time of the same field of view were combined into the same observation if the consecutive time intervals occurred within 24 hours of another clean time interval for the same field of view. Observations with at least 90% of the observing time removed by cleaning cuts or with less than 5 ks left per detector for more than one detector were also removed from further analysis. Based on the spectral testing performed in Ringuette et al. (2021, section 6.1), we also removed observations with total average count rates above 0.065 counts $s^{-1}$ in the 3-7 keV energy range to increase the likelihood of resolving the O VII SWCX emission line in the subsequent spectral analysis.

**Table 1:** Low-Background Observations

| Target Name | Galactic Coordinates (*l, b*) | Start Time (UTC)* | End Time (UTC)* | Sun Angle | Exposure Time (ks) |
|---|---|---|---|---|---|
| MSWCX1 | 126.4°, -44.6° | 11/21/2018 13:35 | 11/22/2018 03:50 | 140.°4 ± 0.°1 | 26.2 |
| | | 10/21/2019 08:40 | 10/22/2019 08:03 | 167.°53 ± 0.°09 | 22.3 |
| | | 01/09/2019 09:24 | 01/09/2019 23:39 | 93.°0 ± 0.°1 | 30.0 |
| MSWCX2 | 152.7°, -54.4° | 11/22/2018 20:26 | 11/23/2018 10:41 | 147.°4 ± 0.°1 | 25.2 |
| | | 10/19/2019 07:32 | 10/20/2019 06:55 | 171.°83 ± 0.°09 | 19.1 |
| MSWCX3 | 163.9°, -46.6° | 11/24/2018 18:42 | 11/25/2018 08:57 | 155.°7 ± 0.°1 | 25.5 |
| | | 08/25/2019 00:49 | 08/26/2019 00:13 | 112.°1 ± 0.°2 | 58.2 |
| | | 10/21/2019 09:03 | 10/22/2019 08:26 | 166.°5 ± 0.°1 | 54.4 |
| | | 11/03/2019 15:48 | 11/04/2019 15:11 | 171.°85 ± 0.°02 | 47.8 |
| | | 08/14/2019 19:00 | 08/15/2019 18:20 | 101.°2 ± 0.°2 | 33.2 |
| MSWCX4 | 156.2°, -28.9° | 10/29/2019 13:27 | 10/30/2019 12:47 | 164.°4 ± 0.°1 | 37.9 |
| | | 10/31/2019 14:33 | 11/01/2019 13:57 | 166.°2 ± 0.°1 | 43.1 |
| | | 08/19/2019 21:54 | 08/21/2019 21:53 | 96.°7 ± 0.°3 | 105.9 |
| MSWCX5 | 218.8°, 28.6° | 02/25/2019 06:20 | 02/25/2019 20:36 | 153.°48 ± 0.°09 | 22.6 |
| | | 01/22/2020 09:41 | 01/23/2020 09:01 | 167.°1 ± 0.°1 | 45.6 |
| | | 02/13/2020 20:44 | 02/14/2020 19:57 | 163.°8 ± 0.°1 | 18.2 |
| | | 11/04/2019 16:46 | 11/05/2019 16:07 | 90.°8 ± 0.°2 | 30.6 |
| MSWCX6 | 199.6°, 42.4° | 01/17/2019 02:43 | 01/17/2019 16:59 | 160.°6 ± 0.°1 | 23.4 |
| | | 02/24/2019 15:19 | 02/25/2019 05:33 | 154.°2 ± 0.°1 | 27.1 |
| | | 03/15/2019 05:29 | 03/15/2019 19:45 | 138.°0 ± 0.°1 | 31.4 |
| | | 02/02/2020 14:49 | 02/03/2020 14:12 | 169.°599 ± 0.°003 | 48.6 |
| MSWCX7 | 211.3°, 50.4° | 02/25/2019 22:07 | 02/26/2019 12:21 | 163.°99 ± 0.°08 | 31.4 |
| | | 03/19/2019 17:20 | 03/20/2019 07:33 | 144.°3 ± 0.°1 | 22.9 |
| | | 02/10/2020 19:24 | 02/11/2020 18:48 | 170.°53 ± 0.°04 | 48.5 |
| | | 03/09/2020 08:19 | 03/10/2020 07:42 | 154.°2 ± 0.°2 | 52.0 |
| | | 03/11/2020 09:19 | 03/12/2020 08:42 | 151.°6 ± 0.°2 | 51.1 |
| | | 05/10/2020 12:41 | 05/12/2020 12:17 | 92.°6 ± 0.°3 | 71.9 |
| MSWCX8 | 241.6°, 45.4° | 02/24/2019 14:56 | 02/25/2019 05:12 | 172.°11 ± 0.°01 | 15.2 |
| | | 03/24/2019 20:34 | 03/25/2019 10:50 | 150.°6 ± 0.°1 | 23.7 |
| | | 03/25/2020 16:01 | 03/26/2020 15:24 | 149.°1 ± 0.°2 | 47.9 |
| | | 05/21/2019 16:51 | 05/22/2019 14:31 | 95.°0 ± 0.°2 | 17.4 |
| MSWCX9 | 216.6°, 61.4° | 02/23/2019 23:55 | 02/24/2019 14:11 | 166.°586 ± 0.°007 | 34.4 |
| | | 03/11/2019 09:05 | 03/11/2019 23:20 | 160.°17 ± 0.°08 | 32.8 |
| | | 03/22/2019 06:58 | 03/22/2019 21:13 | 150.°37 ± 0.°09 | 34.3 |
| | | 03/19/2020 13:12 | 03/20/2020 12:35 | 152.°0 ± 0.°2 | 42.3 |
| | | 05/26/2019 19:54 | 05/27/2019 17:41 | 88.°8 ± 0.°2 | 24.5 |
| | | 05/18/2020 16:09 | 05/22/2020 16:35 | 94.° ± 1.° | 212.7 |
| MSWCX10 | 264.1°, 65.3° | 02/16/2019 22:23 | 02/17/2019 12:40 | 152.°7 ± 0.°1 | 23.2 |
| | | 03/28/2019 16:56 | 03/29/2019 07:12 | 166.°21 ± 0.°09 | 23.4 |
| | | 04/16/2020 01:56 | 04/17/2020 00:54 | 147.°9 ± 0.°2 | 17.2 |
| | | 12/24/2018 23:57 | 12/25/2018 14:13 | 98.°7 ± 0.°1 | 35.0 |
| | | 05/31/2019 22:55 | 06/01/2019 22:18 | 103.°7 ± 0.°2 | 37.8 |
| | | 12/20/2019 17:12 | 12/21/2019 16:35 | 93.°6 ± 0.°2 | 47.7 |
| | | 12/22/2019 18:14 | 12/23/2019 17:37 | 95.°7 ± 0.°2 | 46.7 |
| | | 12/24/2019 19:16 | 12/25/2019 10:41 | 97.°6 ± 0.°1 | 37.1 |
| MSWCX11 | 92.0°, -48.7° | 09/11/2019 10:39 | 09/13/2019 10:39 | 166.°3 ± 0.°1 | 58.4 |
| MSWCX12 | 73.4°, -66.4° | 08/03/2019 12:29 | 08/04/2019 11:52 | 138.°9 ± 0.°2 | 19.6 |
| | | 08/21/2019 23:04 | 08/22/2019 22:28 | 156.°9 ± 0.°2 | 59.1 |
| | | 10/04/2019 02:00 | 10/04/2019 22:18 | 158.°3 ± 0.°1 | 41.4 |
| | | 06/26/2019 14:20 | 06/27/2019 13:43 | 103.°8 ± 0.°2 | 52.4 |
| | | 06/19/2020 05:34 | 06/22/2020 05:33 | 98.°6 ± 0.°4 | 152.7 |

[a]Dates and times given with month, day, and year (MM/DD/YYYY) followed by the time in UTC in hours and minutes.

The surviving observations are presented in Table 1. Columns 1 and 2 give the target name and galactic coordinates for the field of view, followed by the start and end times in UTC for each observation in columns 3 and 4. The last two columns give the average sun angle with the statistical uncertainty and the total exposure time for each observation. Observations of the same field of view are grouped together and then arranged by sun angle range in the table, sun angles greater than 110° first.

Not all attempted observations were successful, whether due to scheduling constraints or high particle backgrounds, preventing us from obtaining observations of all fields of view through the flank of the Earth's magnetosheath. Despite these drawbacks, we obtained a substantial amount of exposure time for our chosen fields of view. After data preparation, we use a total of 1255.9 ks of exposure time of observations down the tail of Earth's magnetosphere (sun angles greater than 110°) and 935.6 ks through its flanks (sun angles less than 110°). The minimum amount of total exposure time for a given field of view regardless of the sun angle is 44.3 ks, ranging up to a maximum of 381.1 ks. Exposure times per field of view are given separately for tail observations in Table 3 and flank observations in Table 4.

### 3. Spectral Modeling

Spectral modeling was performed separately for observations with average sun angles greater than 110°, called 'tail' observations, and for observations with lines of sight closer to the sun, called 'flank' observations, due to the presence of the flanks of the Earth's magnetosheath along the line of sight and the possible m-SWCX contributions from them. All tail observations of the same field of view were combined into a composite spectrum of that field of view to increase the statistics in the analysis. In contrast, all flank observations were analyzed separately to increase the likelihood of detecting the highly temporally varying magnetospheric SWCX emission.

The spectral components in the tail spectra fall into three categories: the slowly varying heliospheric SWCX emission, the invariant astrophysical background (for the time scale of the instrument observation), and the remaining particle background. In addition to these components, the flank spectra possibly contain emission from magnetospheric SWCX. Our approach to modeling these components is described in the following subsections.

*3.1 Modeling Heliospheric SWCX*

Ringuette et al. (2021) tested two methods to calculate the amplitude of the heliospheric SWCX emission in a given observation using a physical model of the expected spectral shape, empirical relations derived from solar wind data, H and He neutral distributions, the solar wind data temporally and spatially relevant for the observation timing and direction, and the location and look direction of the instrument for that observation. The first method assumed a uniformly slow solar wind with constant charge exchange cross sections, while the second method included the variation of solar wind speed and fluxes with heliolatitude. Neither model proved successful for predicting the amount of heliospheric SWCX in the observations taken near the southern ecliptic pole, but the two models' performances were indistinguishably consistent with the observations near the solar ecliptic plane. Since the two methods performed equally well in the ecliptic plane, we take the liberty of choosing the second more physical model and limit our choice of fields of view to those near the ecliptic plane.

The chosen spectral model for the heliospheric SWCX emission present in our observations calculates the spectral shape and amplitudes of the emission. We use the same method to predict the spectral shape of the heliospheric SWCX emission in both the tail and flank spectra with one approximating assumption. In the earlier work, the observations taken were along the tail of the He focusing cone, justifying the assumption in that work of 100% neutral He and negligible amounts of neutral H along those lines of sight. That same assumption is not justified in this work since the look directions of the observations typically do not cross the He-focusing cone (see right panel of Figure 1). The differing contributions of ions resulting from the changing balance between the neutral He and neutral H percentages shifts the position of the averaged emission line (compare the second and third columns in Table 2). However, the maximum difference in SWCX emission line energies was determined in that work to be 6 eV by comparing the line energies resulting from assuming 100% neutral He and 100% neutral H, well below *HaloSat's* energy resolution (Zajczyk et al. 2020). Therefore, we assume the neutral population along the lines of sight to be 50% neutral He and 50% neutral H and simply take the ion line energies calculated in that work as reasonable approximations (see Table 2 below).

Our assumption of the neutral populations along the lines of sight also affects the line ratios reported in Table 2 of that work, specifically the line ratios for the low energy emission line and the secondary O VIII emission line. As described in the previous work, the low energy line is the combination of several ion lines at the low end of *HaloSat's* energy range typically observed as a blended peak (see row 1 of Table 2 and the small peak below 0.5 keV in the SWCX spectral component in Figures 3 and 4). The primary lines are the emission lines of a given ion with the largest intensity (e.g. the OVII a emission line on the second row of Table 2), and the line ratios are the intensity ratios between this primary line and the less intense secondary line of each ion (e.g. 0.126 for OVII b / O VII a in Table 2).

**Table 2** SWCX Spectral Shape

| Ion(s) | Energy (100% H) (keV) | Energy* (100% He) (keV) | Line Ratio (100% H) | Line Ratio (100% He)* | Line Ratio (H = He) |
|---|---|---|---|---|---|
| S IX, N VI, C VI, N VII | 0.4445 | 0.4434 | (Q)/(O VII a) = 0.966 | (Q)/(O VII a) = 0.925 | (Q)/(O VII a) = 0.946 |
| O VII a | 0.5646 | 0.5633 | 1.0 | 1.0 | 1.0 |
| O VII b | 0.6792 | 0.6792 | (Q)/(O VII a) = 0.126 | (Q)/(O VII a) = 0.126 | (Q)/(O VII a) = 0.126 |
| O VIII a | 0.6531 | 0.6531 | 1.0 | 1.0 | 1.0 |
| O VIII b | 0.8081 | 0.8031 | (Q)/(O VIII a) = 0.320 | (Q)/(O VIII a) = 0.549 | (Q)/(O VIII a) = 0.434 |
| Ne IX a | 0.9091 | 0.9087 | 1.0 | 1.0 | 1.0 |
| Ne IX b | 1.1004 | 1.1004 | (Q)/(Ne IX a) = 0.100 | (Q)/(Ne IX a) = 0.100 | (Q)/(Ne IX a) = 0.100 |

*Column repeated from Ringuette et al. (2021). The energies reported in the third column and the line ratios in the last column are used in this work as reasonable approximations.
**Note**: Primary ion lines are indicated with an "a" and secondary lines with a "b." During spectral fitting, the low-energy line is either held fixed relative to O VIIa with the given ratio or fitted simultaneously with the other primary lines.

Since we have assumed an equal balance of neutral H and neutral He along the line of sight, we have assumed that the line ratios are the mean of the ratios expected for a pure H interaction and for a pure He interaction (see the last three columns of Table 2). The difference between these line ratios and those used for this work is 0.1 for the secondary O VII emission, and 0.02 for the low energy line. Given the typical line intensity of a few line units (LU = ph cm$^{-2}$ s$^{-1}$ sr$^{-1}$) for the low energy line observed in the previous work, the resulting line ratio difference of about 0.02 for the low energy emission line results in a difference of emission on the order of the detection uncertainty in the line. For the secondary O VIII emission line, which has a slightly lower typical line intensity of about 1 LU, the line ratio difference results in a difference in line intensity smaller than the observation error reported in that work. Therefore, our choice of line ratio convention does not significantly affect our conclusions.

Table 2 presents the line energies and line ratios used in this work to represent the shape of the heliospheric SWCX contribution in each spectrum (see the third and last column). Given the presumably negligible contribution of magnetospheric SWCX in the tail composite spectra, we fix the calculated amplitudes and line energies of the modeled SWCX emission lines during the spectral fitting process. For the flank spectra, we instead fit for the amplitudes of the primary emission lines including the low energy line, hold the line energy positions and line ratios constant as in the tail spectral analysis process, and compare the observed line intensities to those predicted by the heliospheric SWCX model.

*3.2 Modeling the Astrophysical and Non-X-Ray Components*

Our chosen spectra are of similarly dark fields of view as in Ringuette et al. (2021), are taken by the same instrument, and with an almost identical data preparation process as in that work. So, we take the same approach as Ringuette et al. (2021) for modeling the astrophysical and non-X-ray components of the spectra. The particle background due to local non-X-ray emission was modeled using a power law component with free amplitudes and indices folded through a diagonal response modeled individually for each detector. This allowed the shape and amplitude of the particle background to be fitted independently for each detector and simultaneously with the astrophysical background in both the flank and tail spectra.

Given the lack of bright X-ray sources in our chosen fields of view, the only significant spectral components due to the astrophysical background are the cosmic X-ray background (CXB), the Galactic halo, and the local hot bubble (LHB). We use the same spectral parameters and shapes for the CXB and LHB components as in Ringuette et al. (2021) and assume the astrophysical background is constant for a given field of view over the entire range of observations. We fit for the temperature and amplitude of the Galactic halo component, similarly assuming an absorbed thermal emission model of a collisional ionization equilibrium (CIE) plasma, limiting the fitted temperature range to 0.1-0.35 keV, and fixing the metallicity to the same value. The hydrogen column density for each field of view was calculated in the same manner as LaRocca et al. (2020, section 3.2), who used the Planck optical extinction maps combined with a response-weighted averaging method to determine the equivalent total Galactic column density.

We create the composite tail spectrum for each field of view by combining all observations of a field of view with sun angles greater than 110° into one "tail" spectrum. For each field of view, we analyzed the composite tail spectrum independently to determine the spectral parameters of the astrophysical background in that field of view. We required the Galactic halo parameters to be the same for the three different detectors, but fitted the instrumental particle background independently for each detector. The particle background was expected to vary between the tail and flank observations.

Ringuette et al. (2021) found absorbed higher temperature components in the two fields analyzed. So, we also test each composite tail spectrum for the presence of an absorbed higher temperature component in a similar method. First, we assume the simplest case of a single absorbed temperature component and use the spectral model described above, fitting for the temperature and emission measure independently in each field of view. Then, we hypothesize there are two absorbed temperature components of distinct temperatures in each spectrum. We fit for the temperatures and emission measures of both components simultaneously, limiting the lower temperature component to a temperature range of 0.1-0.35 keV as described before, and limiting the higher temperature component to a temperature range of 0.6-1.5 keV to prevent spectral confusion between the fitted temperatures as in Ringuette et al. (2021). If the fitted emission measure of the higher temperature component is within three standard deviations of zero, we disregard the higher temperature hypothesis and conclude only a single temperature component is significant in that field of view.

The final parameters of the astrophysical background determined in this manner using the composite tail spectrum in each field of view are then held constant during the spectral analysis of the individual flank spectra for the same field of view. In the analysis of the flank spectra, we instead fit for the line intensities of the primary and low energy SWCX emission lines (Table 2), holding the line ratios and energies constant as in Ringuette et al. (2021), and simultaneously fitting for the shapes and amplitudes of the particle background on each detector.

**4. Spectral Analysis Results**

*4.1 Galactic Halo*

We report the final parameters for the Galactic halo component in each of the twelve composite tail spectra analyzed in Table 3, including the calculated Galactic column densities for each field of view. The target name, Galactic coordinates, and exposure time for each field of view are given in the first three columns, followed by the calculated absorption in the fourth column. The fitted temperature and emission measure of the Galactic halo component in each field are given in the fifth and sixth columns with the fitted 90% confidence intervals, followed by the number of degrees of freedom and the reduced $\chi^2$ for each spectral fit in the seventh and penultimate columns. The last column gives the number of flank spectra obtained for each field of view.

As indicated in the table, spectral testing of the MSWCX3 and MSWCX6 fields of view concluded a significant higher temperature component to be present in both fields of view. The temperature and emission measure of the higher temperature components in these two fields of view are given in the table footnote, along with the significance of each component and the related null probabilities. The final spectral fits for these two composite tail spectra are presented in Figures 3 and 4.

Ringuette et al. (2021) considered alternate contributions, such as emission from T Tauri or M Dwarf stars, for the high temperature components reported in that work. Here, we must reject the hypothesis of any possible contribution from these types of stars in both the MSWCX3 and MSWCX6 fields of view since they are both more than 30° from the Galactic plane, where T Tauri stars and M Dwarf stars are populous. We also find that the fitted temperatures of the second thermal component are consistent with the value reported in Das et al. (2019b): $1.09^{+1.9}_{-0.64}$ keV (90% CI) based on the Ne X absorption line, but not with the temperatures reported in Das et al. (2019a) for emission. Following the same argument in the Ringuette et al. (2021) work, we also conclude that the small-amplitude high-temperature emission could possibly be from diffuse gas of ≈$10^7$ K in the Galactic halo along the line of sight.

*4.2 SWCX*

The fitted parameters in Table 3 were held fixed during the spectral analysis of the flank spectra obtained for the same field of view, while the best fit line intensities for the primary and low energy lines were obtained. The final line intensities for the most significant line, O VII, are given in Table 4 and compared to those predicted by the heliospheric SWCX model. The first three columns give the target name, observation start date in UTC, and cleaned exposure time for each flank observation obtained. Column 4 presents the best fit line intensity for the O VII line, followed by the predicted h-SWCX O VII line intensity and the difference between them, each with 1σ errors. The final two columns show the number of degrees of freedom and the reduced $\chi^2$ for each spectral fit.

**Table 3:** Galactic Halo Results

| Target Name | Galactic Coordinates (*l, b*) | Tail Exp. Time (ks) | $n_H$ ($10^{20}$/cm$^2$) | Warm-Hot Temperature (keV) | Warm-Hot Emission Measure^ | DoF | $X_\nu^2$ | # Flank Obs. |
|---|---|---|---|---|---|---|---|---|
| MSWCX1 | 126.4°, -44.6° | 48.6 | 3.63 | 0.212±0.013 | 14.3±2.4 | 179 | 1.091 | 1 |
| MSWCX2 | 152.7°, -54.4° | 44.3 | 2.65 | $0.210^{+0.016}_{-0.015}$ | $12.6^{+2.2}_{-2.3}$ | 158 | 0.903 | 0 |
| MSWCX3* | 163.9°, -46.6° | 185.9 | 6.94 | $0.194^{+0.026}_{-0.024}$ | $6.8^{+1.5}_{-1.3}$ | 441 | 1.014 | 1 |
|  |  |  |  | $0.73^{+0.12}_{-0.10}$ | $1.65^{+0.33}_{-0.55}$ |  |  |  |
| MSWCX4 | 156.2°, -28.9° | 81.0 | 13.33 | $0.238^{+0.024}_{-0.019}$ | $9.9^{+1.9}_{-2.2}$ | 271 | 1.110 | 1 |
| MSWCX5 | 218.8°, 28.6° | 86.4 | 3.04 | $0.223^{+0.020}_{-0.018}$ | 7.1±1.6 | 295 | 1.024 | 1 |
| MSWCX6* | 199.6°, 42.4° | 130.6 | 1.94 | $0.181^{+0.0083}_{-0.019}$ | 9.2±1.2 | 421 | 1.103 | 0 |
|  |  |  |  | $0.85^{+0.20}_{-0.16}$ | $0.96^{+0.31}_{-0.51}$ |  |  |  |
| MSWCX7 | 211.3°, 50.4° | 206.0 | 2.09 | $0.201^{+0.010}_{-0.009}$ | 9.5±1.1 | 498 | 1.062 | 1 |
| MSWCX8 | 241.6°, 45.4° | 86.8 | 2.19 | 0.202±0.016 | 8.4±1.6 | 282 | 1.095 | 1 |
| MSWCX9 | 216.6°, 61.4° | 143.9 | 1.10 | 0.197±0.010 | $10.1^{+1.2}_{-1.3}$ | 414 | 1.125 | 2 |
| MSWCX10 | 264.1°, 65.3° | 63.9 | 1.76 | 0.204±0.015 | $10.2^{+2.0}_{-1.9}$ | 270 | 0.941 | 5 |
| MSWCX11 | 92.0°, -48.7° | 58.4 | 4.90 | $0.193^{+0.023}_{-0.024}$ | $8.5^{+2.6}_{-2.4}$ | 231 | 0.987 | 0 |
| MSWCX12 | 73.4°, -66.4° | 120.1 | 1.30 | 0.208±0.011 | 10.4±1.3 | 351 | 1.190 | 2 |

^Emission measure in units of x10$^{-3}$ cm$^{-6}$ pc. All errors are 90% confidence intervals.
*These two spectra were best fit with a double thermal component for the Galactic Halo emission. The temperature and emission measure for the second thermal component for MSWCX3 (MSWCX6) are given below the Warm-Hot parameters for each of the two targets in the table and were detected at a significance of 4.97σ (3.09σ) with F-test probabilities of 2.65x10$^{-5}$ (1.69x10$^{-2}$) comparing double to single thermal components.

Figure 5 provides a more visual comparison of the fitted and predicted O VII line intensities. The fitted O VII line intensities are plotted against the predicted line intensities with error bars representing the 1σ errors given in Table 4. The solid line indicates the null hypothesis of the best fit line intensities being equal to those predicted for a given observation. Values plotted above this line indicate observed emissions greater than the predicted emissions, and values plotted below this line show an observation with lower observed emission than predicted by the model. Out of the 15 flank observations obtained, four are noted to have fitted O VII line intensities more than 3σ above the predicted value for that observation (see footnote of Table 4). The observation with an O VII line intensity excess greater than 4σ is indicated in Figure 5 with a red square, and the remaining three observations with excesses above 3σ but below 4σ are indicated with yellow diamonds. These four observations are further analyzed in the next section.

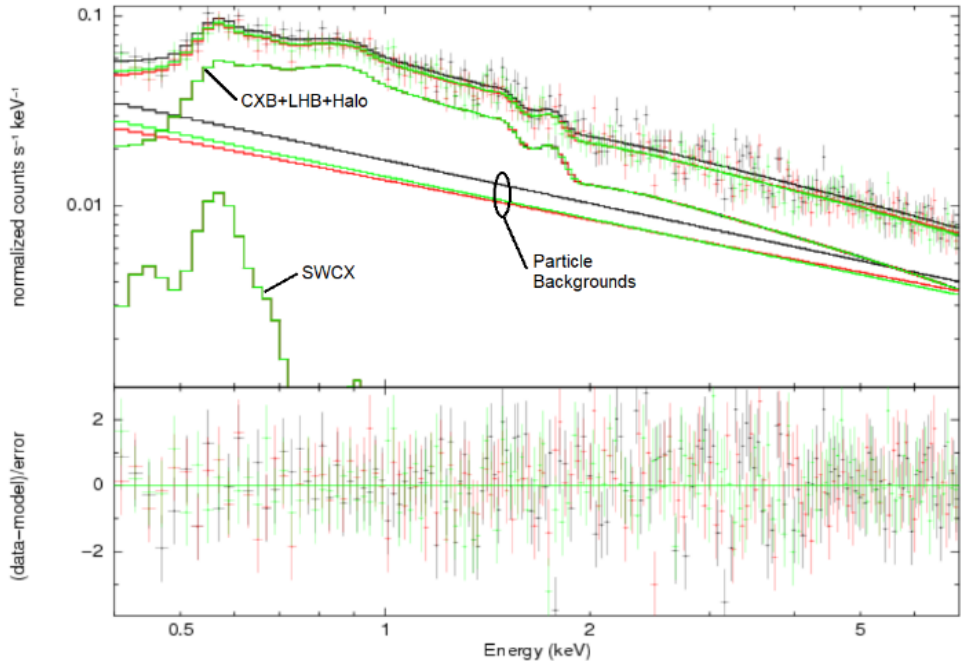

**Figure 3:** Spectral fit for MSWCX3 tail composite spectrum. Solid curves show the spectral components as labeled: heliospheric SWCX emission (SWCX), emission from the CXB, LHB, and a two-temperature Galactic Halo (CXB+LHB+Halo), the three particle backgrounds (straight lines; fitted), and the total spectral model at the top. Crosses show the data and residuals are shown in the bottom panel. The Ne IX SWCX contributions are barely visible just below 1 keV.

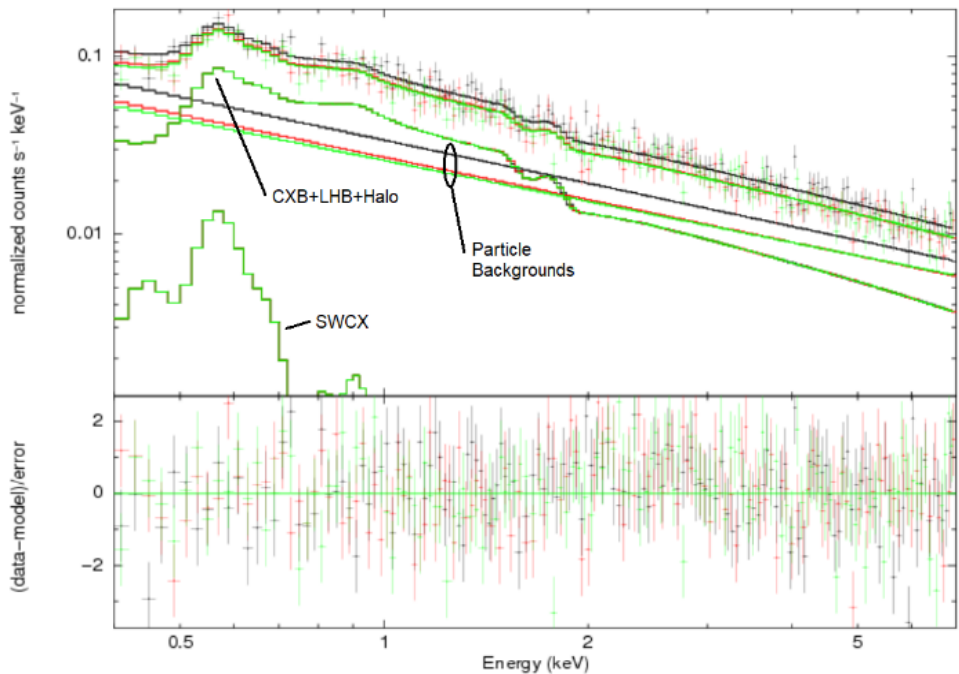

**Figure 4:** Spectral fit for MSWCX6 tail composite spectrum. Solid curves show the spectral components as labeled: heliospheric SWCX emission (SWCX), emission from the CXB, LHB, and two-temperature Galactic Halo (CXB+LHB+Halo), the three particle backgrounds (straight lines; fitted), and the total spectral model at the top. Crosses show the data and residuals are shown in the bottom panel. The Ne IX SWCX contributions are barely visible just below 1 keV.

**Table 4:** SWCX Results

| Target Name | Start Date (MM/DD/YYYY hh:mm) (UTC) | Exposure Time (ks) | m-SWCX + h-SWCX Observed O VII (LU) | Predicted h-SWCX O VII (LU) | Difference (Obs-Exp) (LU) | DoF | $\chi^2_\nu$ |
|---|---|---|---|---|---|---|---|
| MSWCX1 | 01/09/2019 09:24 | 30.0 | $0.39^{+0.59}_{-0.24}$ | 1.135±0.048 | $-0.74^{+0.59}_{-0.24}$ | 130 | 1.014 |
| MSWCX3* | 08/14/2019 19:00 | 33.2 | $2.05^{+0.44}_{-0.41}$ | 0.707±0.036 | $1.35^{+0.44}_{-0.41}$* | 115 | 1.042 |
| MSWCX4* | 08/19/2019 21:54 | 105.9 | 1.78±0.25 | 0.727±0.032 | $1.06^{+0.25}_{-0.26}$* | 263 | 1.028 |
| MSWCX5 | 11/04/2019 16:46 | 30.6 | $2.47^{+0.58}_{-0.57}$ | 1.658±0.050 | $0.82^{+0.58}_{-0.57}$ | 106 | 0.894 |
| MSWCX7 | 05/10/2020 12:41 | 71.9 | 1.53±0.39 | 0.902±0.040 | 0.62±0.39 | 218 | 0.981 |
| MSWCX8* | 05/21/2019 16:51 | 17.4 | $4.09^{+0.84}_{-0.72}$ | 1.470±0.075 | $2.62^{+0.84}_{-0.72}$* | 70 | 0.919 |
| MSWCX9 | 05/26/2019 19:54 | 24.5 | $2.72^{+0.67}_{-0.68}$ | 1.477±0.069 | 1.24±0.68 | 103 | 1.004 |
| MSWCX9 | 05/18/2020 16:09 | 212.7 | $1.59^{+0.23}_{-0.22}$ | 0.937±0.044 | 0.65±0.23 | 448 | 1.149 |
| MSWCX10 | 12/24/2018 23:57 | 35.0 | $2.84^{+0.58}_{-0.60}$ | 1.050±0.047 | $1.79^{+0.59}_{-0.60}$ | 109 | 0.824 |
| MSWCX10* | 05/31/2019 22:55 | 37.8 | $3.20^{+0.49}_{-0.50}$ | 1.444±0.077 | $1.75^{+0.50}_{-0.51}$* | 172 | 1.350 |
| MSWCX10 | 12/20/2019 17:12 | 47.7 | $2.11^{+0.50}_{-0.49}$ | 1.087±0.069 | $1.03^{+0.50}_{-0.49}$ | 199 | 1.016 |
| MSWCX10 | 12/22/2019 18:14 | 46.7 | 1.48±0.49 | 1.125±0.061 | 0.35±0.49 | 170 | 1.035 |
| MSWCX10 | 12/24/2019 19:16 | 37.1 | $1.52^{+0.53}_{-0.54}$ | 0.984±0.058 | 0.53±0.54 | 126 | 1.326 |
| MSWCX12 | 06/26/2019 14:20 | 52.4 | $0.67^{+0.43}_{-0.41}$ | 1.497±0.058 | $-0.82^{+0.44}_{-0.41}$ | 183 | 1.257 |
| MSWCX12 | 06/19/2020 05:34 | 152.7 | $0.64^{+0.25}_{-0.26}$ | 1.107±0.044 | -0.47±0.26 | 371 | 1.077 |

All reported errors are 1σ errors. The errors on the observed line intensities were calculated from the fitted 90% confidence intervals from Xspec.

*The significance of the observed value above the expected value is greater than 3σ for these observations (3.31σ, 4.12σ, 3.62σ, and 3.45σ, respectively).

**Section 5: The Magnetospheric SWCX Component**

The m-SWCX component in each of the four observations highlighted in Figure 5 was calculated following the method described in section 6.2.1 of Ringuette et al. (2021). The O VII emission due to the m-SWCX contribution was calculated by integrating the product of the ion flux, the neutral density, and the charge exchange cross section along the line of sight according to Equation 1 below (as done in section 6.2.1 of Ringuette et al. 2021). The proton density and velocity is again highly variable along this path, so was extracted from BATS-R-US simulation runs (Toth et al. 2005). The input values to the simulation runs were chosen to reflect the solar wind conditions during the observing times included in the cleaned observations. The four simulation run outputs are accessible on the CCMC website (see the Acknowledgement section for more details).

$$I = \frac{1}{4\pi} \int_{obs}^{\infty} F_{O+7}(s) N_H(s) \sigma_{H,O+7} Y_{O\ VII,H} ds$$
$$= \frac{1}{4\pi} \frac{N_{O+7}}{N_P} \int_{obs}^{\infty} N_P(s) V_P(s) N_H(s) \sigma_{H,O+7} Y_{O\ VII,H} ds.$$

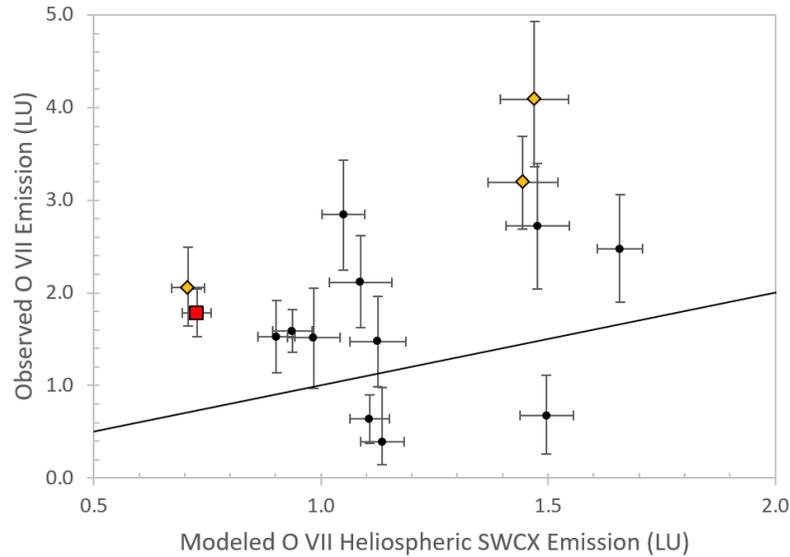

**Figure 5:** Observed vs Predicted O VII Line Intensities for MSWCX flank observations. Vertical error bars indicate 1 sigma error bars calculated using the fitted 90% confidence intervals, horizontal error bars indicate 1 sigma errors on predicted values. The solid line shows the hypothesis of observed O VII line intensities equal to the predicted values. Markers above and inconsistent with this line are indicative of excess emission, optimistically due to magnetospheric SWCX emission long the line of sight. Yellow diamonds (red square) indicate observations with at least 3σ (4σ) excess emission.

The calculated O VII line intensities for the asterisked MSWCX3, MSWCX4, MSWCX8, and MSWCX10 observations in Table 4 are 0.107 LU, 0.044 LU, 0.053 LU, and 0.055 LU. Notably, these values are all an order of magnitude lower than the observed O VII excesses. We find that 1.02-2.57 LU are not accounted for in the four highlighted observations. Since the observations were conducted in the same manner as those analyzed in the 2021 paper, the same calculations in that paper concerning atmospheric contamination applies to these observations. In short, no significant atmospheric contamination is found to significantly contribute at the O VII energy.

We are therefore left to conclude one or more of the following: (1) the solar wind parameters used in the calculations do not properly represent the solar conditions along the observational lines of sight through the emissive portions of the Earth's magnetosphere, (2) the BATS-R-US model output produced does not accurately simulate the proton density and velocity in the regions sampled, and (3) there are other possible contributions to the O VII line energy band that we do not yet understand.

Our previous work on modeling the h-SWCX emission in the ecliptic plane showed that currently available observational solar wind data is a reasonable approximation of the long-term averaged solar wind behavior along lines of sight away from the emissive portions of the Earth's magnetosphere, as is relevant for h-SWCX emission calculations. However, the highly variable behavior of the solar wind, both temporally and spatially, can result in dramatic differences between the solar wind conditions at the instrument and near the Earth as demonstrated in difficulty of forecasting space weather based on data observed at L1. Consider the oft occurring situation of an oblique shock front propagating through the L1 location. The enhancement in the solar wind in the passing front is registered by the solar wind instruments near that location, but the shock front never reaches the Earth due to the direction of its overall velocity vector, so the forecasted solar wind enhancement and subsequent geomagnetic storm effects are not observed at Earth. It is these and similar known differences between the solar wind data and the observed effects near the Earth that are likely a cause of a portion of the O VII emission not accounted for.

Concerning the accuracy of the BATS-R-US model close to the Earth, we acknowledge our probable lack of understanding on the best ways to use the model. We propose that there are improvements to be made in this area, and look to collaborate with geospace researchers on this matter in a future work. To our knowledge, this work and our previous work are the first validation efforts of the BATS-R-US model for the proton density and velocity near the Earth. As such, it is common that the match between observational and simulated data is qualitatively reasonable, but quantitatively poor. We also note that the modeled predictions do not give any uncertainty in the predicted values, and that the calculation of uncertainties in modeled data is a topic of current research. We applaud the significant efforts of the BATS-R-US modeling team and their contribution to the field, and invite them to use this work to investigate our own errors in our modeling efforts, and what other causes there may be of the differences between these data and the model results. Perhaps this and similar works could propel the long-standing collaboration between astrophysics and heliophysics into a new territory: model validation studies of heliophysics models using astrophysics SWCX data.

**Section 6: Comparison of the Galactic Halo Results**

As a secondary result, this work reports two fields of view with non-negligible absorbed higher temperature components, specifically MSWCX3 and MSWCX6 with hot temperatures of $0.73^{+0.12}_{-0.10}$ and $0.85^{+0.20}_{-0.16}$ keV and corresponding emission measures with significances of 4.97σ and 3.09σ. Ringuette et al. (2021) also reported non-negligible absorbed high temperature emission with a slightly higher temperature of ~1 keV for both fields of view analyzed using the same spectral analysis methods. Additionally, Bluem et al. (2022a) reported several fields of view with significant absorbed higher-temperature components. The 'hot' temperatures reported there are typically lower but do not exclude the corresponding values reported here. The values reported here for the lower temperature components range from ~0.18 to ~0.24 keV, with the largest two values occurring for the fields of view within 30° of the Galactic equator. Excluding those fields as done in the Bluem et al. (2022a) work results in a temperature range of ~0.18 to ~0.21 keV, which is not consistent with the stacked fit temperature reported in the Bluem et al. (2022a) work, but are included in the range of individual fitted temperatures reported there.

Figures 6 and 7 compare the results per field in more detail using the results reported here and in the VizieR catalog associated with the mentioned paper (Bluem et al. 2022b). The figures present the fitted temperatures (top panels) and emission measures (bottom panels) for both the warm (Figure 6) and hot (Figure 7) components of the absorbed emission. One main difference in the two analyses is that this work reports a hot temperature and emission measure *only if* the fitted hot emission measure was at least 3σ above zero; otherwise, a single temperature model was assumed. In contrast, the Bluem et al. (2022a) analysis includes the hot temperature component in the spectral model for all fields regardless of the component's significance. Also, the Bluem works exclude fields of view within 30° of the Galactic plane (e.g. MSWCX4 and MSWCX5, aka Vizier ID numbers 71 and 72, see Table 5). Despite the differing analysis methods, the temperatures and emission measures are generally consistent for the two fields where this analysis detected a significant hot emission measure and retained that component in the spectral model. Otherwise, the temperatures of the warm component reported in Bluem et al. (2022b) are typically lower than those reported here. The two fields analyzed solely in this paper have a warm emission measure in the same range as the other emission measures and a slightly higher but not inconsistent warm temperature than the other temperatures reported here.

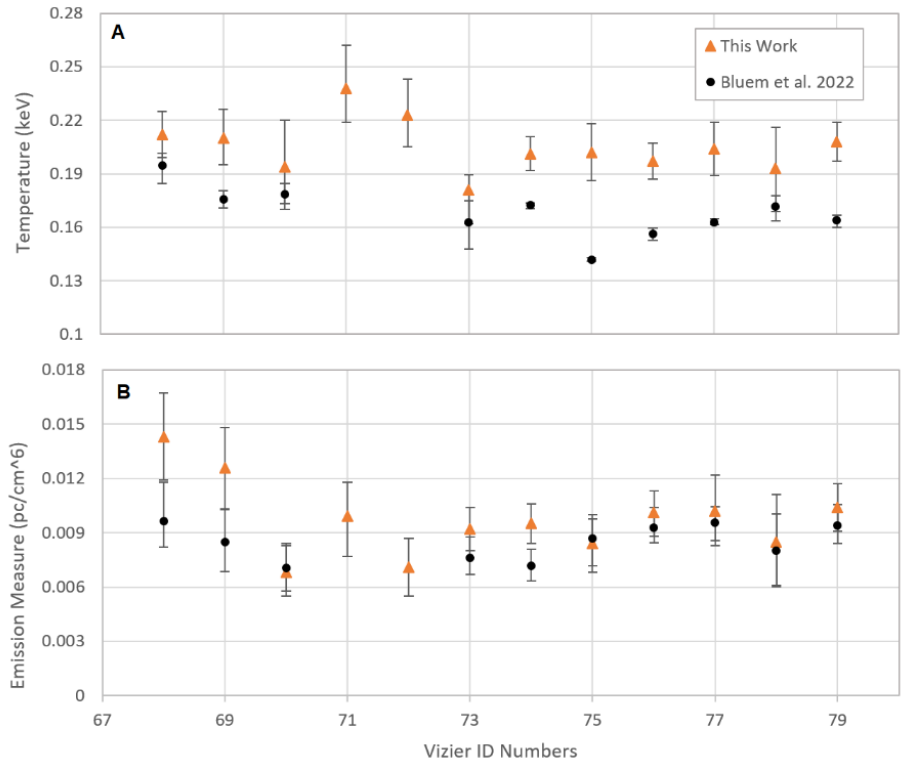

**Figure 6:** Comparison of the Temperature and Emission Measure of the Warm component. Panel A (top) compares the fitted temperatures in keV and panel B (bottom) compares the emission measures of the warm components of the Galactic Halo measured in this work (orange triangles) and the Bluem et al. (2022b) work (black circles). The data were obtained from the Vizier catalog in Bluem et al. (2022b) with the identification numbers noted on the horizontal axis (67 = MSWCX1, 68 = MSWCX2, etc., see Table 5). All error bars represent 90% CIs. The emission measures are generally consistent, but the temperatures are shifted lower in that work.

The residuals due to the gain shift and edge effect reported in the Bluem et al. (2022a) work are not observed here (see Figures 3 and 4), possibly due to the lower statistics from the lower exposure times (see Appendix A of Bluem et al. (2022a)). Unfortunately, the new analysis methods were not available in time to apply to this work. While it would be interesting to reanalyze these fields with the gain shift and edge correction incorporated, it is beyond the scope of this work.

The warm component temperatures reported in this work are typically higher than those reported in Bluem et al. (2022a) and the associated Vizier catalog (Bluem et al. 2022b) due to the differing analysis methods. One such difference is the more complete representation of SWCX emission lines in this work. Excluding the collection of low energy SWCX emission lines in that work may have motivated the addition of the lower energy background power law. Those photon indices were linked across detectors just as the astrophysical parameters were, possibly signaling a link to a non-local cause (e.g. low energy SWCX emission along the line of sight). Similarly excluding the higher energy SWCX emission lines could have improperly but slightly enhanced the emission measure of the higher temperature component, several of which are barely above $3\sigma$ (see Table 5 for a sample). As a result, more of the detected hot components in the related Vizier catalog likely have a true significance lower than $3\sigma$.

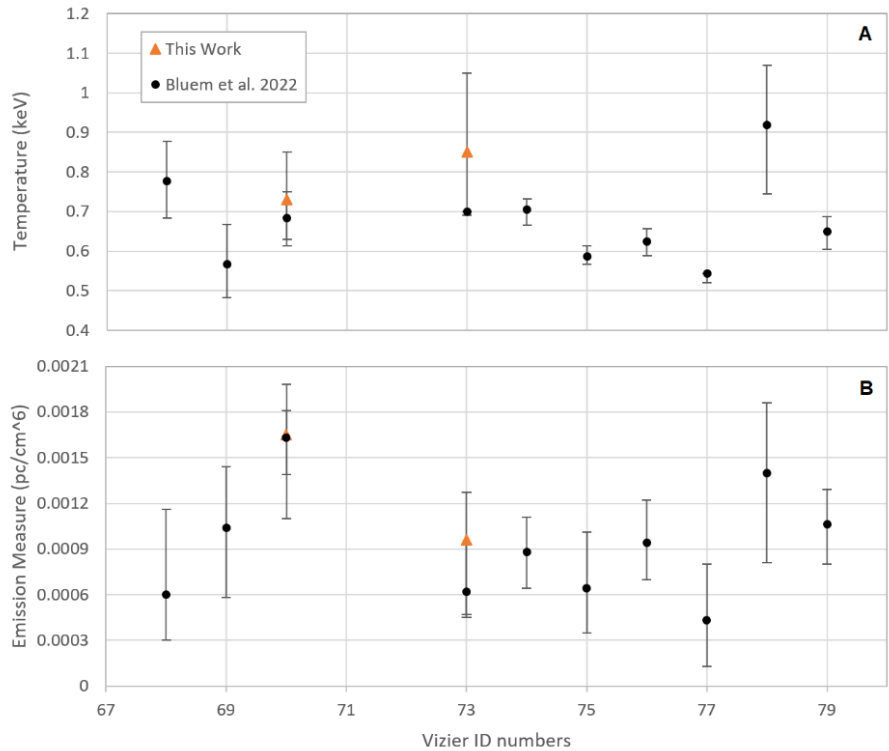

**Figure 7:** Comparison of the Temperature and Emission Measure of the Hot component. Panel A (top) compares the fitted temperatures in keV and panel B (bottom) compares the emission measures of the hot components of the Galactic Halo measured in this work (orange triangles) and the Bluem et al. (2022b) work (black circles). The data were obtained from the Vizier catalog in Bluem et al. (2022b) with the identification numbers noted on the horizontal axis (67 = MSWCX1, 68 = MSWCX2, etc). All error bars represent 90% CIs. The emission measures and temperatures are consistent for the two fields measured by both works. Note that a hot temperature component is only reported from this work if the emission measure is more than three standard deviations above zero, where one standard deviation is defined as the lower limit of the confidence interval divided by the ratio 1.645.

**Table 5:** ID Numbers and Exposures of the Compared Fields

| Field Name (this work) | Vizier ID Number | Exposure Time (this work, ks) | Exposure Time (Vizier, ks) | Significance of Hot Emission Measure (Vizier) |
|---|---|---|---|---|
| MSWCX1 | hs0068 | 48.6 | 78.0 | 3.29 |
| MSWCX2 | hs0069 | 44.3 | 76.4 | 3.72 |
| MSWCX3 | hs0070 | 185.9 | 192.3 | 11.17 |
| MSWCX4 | hs0071 | 81.0 | --- | --- |
| MSWCX5 | hs0072 | 86.4 | --- | --- |
| MSWCX6 | hs0073 | 130.6 | 165.7 | 6.80 |
| MSWCX7 | hs0074 | 206.0 | 258.2 | 6.03 |
| MSWCX8 | hs0075 | 86.8 | 161.7 | 3.63 |
| MSWCX9 | hs0076 | 143.9 | 161.0 | 6.44 |
| MSWCX10 | hs0077 | 63.9 | 191.8 | 2.36 |
| MSWCX11 | hs0078 | 58.4 | 58.4 | 3.90 |
| MSWCX12 | hs0079 | 120.1 | 162.2 | 6.71 |

Note that several of the hot emission measures reported in Bluem et al. (2022) are below 4σ and one is below 3σ despite the typically much larger exposure times than used in this work. The two fields reported in this work with a significant absorbed hot temperature component have the highest significances in the Vizier database.

At the time of writing, the Vizier catalog also contains spectral results for hot components with emission measures within 3σ of zero, which would have been deemed negligible and replaced with a single temperature component in the approach presented here (see MSWCX10 in Table 5). Including these insignificant higher temperature components as a standard feature in a spectral fit decreases the fitted temperature of the warm component and assumes that the higher temperature component is universal yet clumpy as found for the warm temperature emission (Kaaret et al. 2020), as would be expected if the emission were associated with the Galactic Halo. In the case of the higher temperature component, this 'clumpiness' includes the possibility of gaps in the cloud of material producing the emission (e.g. an emission measure consistent with zero), conceptually similar to a partly- or mostly-cloudy sky in terrestrial weather. However, excluding those components as done in this work assumes that such higher temperature emission is not universal, which would be correct if the emission is not associated with the Galactic Halo. The work done here primarily focuses on finding a reasonable model for the astrophysical emission as one step towards separating emission generated by magnetospheric SWCX from the other sources, not debating what the source of one of the components is. We choose the second assumption and only report higher temperature components for which a significant emission measure is determined. However, we include the spreadsheets containing the information for our spectral fits for all fields with both the warm and hot components on the associated GitHub repository for the interested reader (Ringuette 2023).

Further observations and work remain to be done on this topic, particularly observations with higher spectral resolution to distinguish the source of the emission by finely resolving the emission lines. In the meantime, it would be interesting for the Bluem et al. 2022a work to be repeated with a more complete SWCX representation such as the one used here. To investigate a possible scenario where the emission is not associated with the Galactic Halo, one could remove all higher temperature components with a significance less than 3σ in the reanalyzed work and change those to instead be modeled with a single temperature model. One could reperform the stacked analysis done there separately for both the spectra with a significant higher temperature component and spectra without such a component to better understand the different scenarios.

Also, it is curious that there is a signature at 3 keV in the stacked CGM spectrum compared to the lack of such a signature in both the Crab and CasA spectra (compare Figures 11, 13 and 14b in that work) and the ECL stacked spectrum (Figure 8 of Ringuette et al. (2021)), even though those spectra have good statistics and were created with similar count rate cuts on the same high energy band (0.25 c/s, 0.25 c/s, and 0.50 c/s on the 3 – 7 keV energy band) (see also the related note in Bluem et al. 2023). Once the spectral modeling is refined as suggested here, one could further investigate the signature at 3 keV, namely to investigate a possible faint signature of higher energy emission in the stacked CGM spectrum at higher Galactic latitudes. One possible path for this investigation would be to raise the count rate cut value until the edge disappears and compare spectral fit results for a variety of physically motivated spectral models (e.g. lingering higher energy emission from a previously active Galactic jet). Detection of a temperature component higher than those discussed in this and the Bluem et al. (2022) paper could simply be indicative of a wider range of Galactic Halo temperatures than currently modeled as posited in that paper.

**Section 7: Conclusions**

We primarily report significant detections of magnetospheric SWCX in four observations. Our observational strategy separated the magnetospheric SWCX emission from the similarly shaped emission of heliospheric SWCX and the warm-hot Galactic halo contributions. These observations were planned as part of the secondary science goal of *HaloSat* - to study SWCX. We successfully obtained clean observations of dark portions of the sky with lines of sight away from the emissive portions of the Earth's magnetosheath (tail observations) and clean observations of the same fields of view with lines of sight through more emissive portions of the Earth's magnetosheath (flank observations). By using the tail observations and our previously validated model of heliospheric SWCX emission, we were able to determine the excess O VII emission in four flank observations (1.06 – 2.62 LU) as compared to the astrophysical background and the expected contribution from the heliospheric SWCX along the line of sight.

The excess O VII emission observed is not explained by atmospheric contamination or by the simulation results obtained for the magnetospheric SWCX contribution to the O VII emission line. We therefore conclude the excess emission to be a result of the spatial and temporal variability of the solar wind, causing poor predictions and simulations of magnetospheric SWCX contributions near the Earth. It is also possible that we have not used the model in the manner most relevant to the situation. Additionally, it is also quite possible that there are other sources of O VII emission along the line of sight that we do not yet understand. We recommend a collaboration between astrophysicists and space weather modelers as the two fields mature further to improve modeling efforts through similar model validation studies.

Secondly, we report a non-negligible (4.97σ and 3.09σ) absorbed ~1 keV high-temperature component in two out of the twelve fields of view analyzed near the ecliptic plane. Both fields of view are far from the ecliptic plane, precluding T Tauri and M Dwarf stars as possible emission sources for this component. The high-temperature emission could be from diffuse gas with a temperature near $10^7$ K in the Galactic Halo along the line of sight.

Improved results for magnetospheric SWCX detection can be obtained with additional X-ray observations of the same fields of view with lines of sight through the emissive portions of the Earth's magnetosheath near solar maximum, when the desired emission enhancement typically occurs more often due to the enhanced solar activity. Additionally, the detection of high temperature components can be further investigated with additional X-ray observations of the same fields of view and an updated analysis procedure as described in Section 6.

The processing files used in this analysis are available online through HEASARC's archive and on the related GitHub repository, which includes the scripts used in this analysis (Ringuette 2023). As of the date of this writing, all data from the HaloSat mission are available through the same HEASARC archive. The HaloSat CubeSat reentered the atmosphere in 2021 January.

**Acknowledgements**: All spectral analyses were performed using XSPEC v12.11.1 (Arnaud 1996), which minimizes the differences between the data and the total model by changing the free parameters until the best fit is found. The spectral analysis was performed with the original calibration files, specifically the arf and rmfs posted on the related GitHub repository (Ringuette 2023). All python codes used in this analysis were written in Python 2.7.15 and are available at the same GitHub repository. The authors acknowledge the use of SWICS data provided by the ACE Science Center (http://www.srl.caltech.edu/ACE/ASC/) and especially wish to extend their thanks to Jim Raines for endless discussions on the data. The authors also acknowledge the use of OMNI data provided by NASA/GSFC's Space Physics Data Facility's OMNIWeb service and magnetospheric simulation results from the Community Coordinated Modeling Center at Goddard Space Flight Center. The CCMC is a multiagency partnership between NASA, AFMC, AFOSR, AFRL, AFWA, NOAA, NSF, and ONR. The SWMF/BATS-RUS model was developed by Tamas Gombosi et al. at the Center for Space Environment Modeling, University of Michigan. The model runs used in this work are named *KD_Kuntz_071921_1, KD_Kuntz_071921_2, KD_Kuntz_071921_3,* and *KD_Kuntz_071921_4* and are available by request through the CCMC website. The full metadata record for these runs can be obtained at (https://ccmc.gsfc.nasa.gov/ungrouped/GM_IM/GM_main.php). The HaloSat mission was supported by NASA grant NNX15AU57G. A portion of R.R.'s work was supported by ADNET Systems, Inc., and D.K.'s modeling work was supported by CNES and performed with the High Performance Computer and Visualisation platform (HPCaVe) hosted by UPMC-Sorbonne Université.

**ORCID IDs:**
R. Ringuette: https://orcid.org/0000-0003-0875-2023
K. D. Kuntz: https://orcid.org/0000-0001-6654-5378
D. Koutroumpa: https://orcid.org/0000-0002-5716-3412
P. Kaaret: https://orcid.org/0000-0002-3638-0637
D. LaRocca: https://orcid.org/0000-0002-7529-4619
J. Richardson: https://orcid.org/0000-0001-8138-7582